**Brain Modularity Mediates the Relation between Task Complexity and Performance**

Qiuhai Yue[a], Randi Martin[a], Simon Fischer-Baum[a], Aurora I. Ramos-Nuñez[a], Fengdan Ye[b,d], and Michael W. Deem[b,c,d]

[a] Department of Psychology, Rice University

[b] Department of Physics & Astronomy, Rice University

[c] Department of Bioengineering, Rice University

[d] Center for Theoretical Biological Physics, Rice University

Corresponding author
Qiuhai Yue
Department of Psychology, MS-25
Rice University
P.O. Box 1892
Houston, TX 77251

Tel: 713-348-3039
qy6@rice.edu

**Abstract**

Recent work in cognitive neuroscience has focused on analyzing the brain as a network, rather than as a collection of independent regions. Prior studies taking this approach have found that individual differences in the degree of modularity of the brain network relate to performance on cognitive tasks. However, inconsistent results concerning the direction of this relationship have been obtained, with some tasks showing better performance as modularity increases and other tasks showing worse performance. A recent theoretical model (Chen & Deem, 2015) suggests that these inconsistencies may be explained on the grounds that high-modularity networks favor performance on simple tasks whereas low-modularity networks favor performance on more complex tasks. The current study tests these predictions by relating modularity from resting-state fMRI to performance on a set of simple and complex behavioral tasks. Complex and simple tasks were defined on the basis of whether they did or did not draw on executive attention. Consistent with predictions, we found a negative correlation between individuals' modularity and their performance on a composite measure combining scores from the complex tasks but a positive correlation with performance on a composite measure combining scores from the simple tasks. These results and theory presented here provide a framework for linking measures of whole brain organization from network neuroscience to cognitive processing.

Keywords: modularity, resting-state fMRI, task complexity, individual difference, network neuroscience

**Brain Modularity Mediates the Relation between Task Complexity and Performance**

Recent work in cognitive neuroscience has moved away from analyzing the function of localized brain structures towards analysis of the brain as a network. Previous studies have shown that the human brain is organized into different subnetworks, termed modules (He et al., 2009; Meunier, Lambiotte, & Bullmore, 2010; Meunier, Lambiotte, Fornito, Ersche, & Bullmore, 2009b; Power et al., 2011; Sporns & Betzel, 2016), which support different aspects of behavioral function, e.g., modules for visual perception, motor control, and attentional processing (Bassett et al., 2011; Bassett, Yang, Wymbs, & Grafton, 2015; Bertolero, Yeo, & D'Esposito, 2015; Smith et al., 2009). Understanding the network organization of the brain has been advanced using graph theory to quantify various local and global properties of its organization, with the nodes of the network defined by brain regions, and edges between nodes as connections, determined either by structural connectivity derived from diffusion tensor imaging or by functional connectivity derived from resting-state or task-based fMRI, between these regions (Bullmore & Sporns, 2009; Finn et al., 2015; Rosenberg et al., 2016; Rubinov & Sporns, 2010). The current work focuses on a global measure of network organization, modularity. Modularity characterizes the community structure of the network in terms of the relative strength of within-module connections and between-module connections. Using Newman's algorithm, nodes in the network are sorted into modules so as to maximize this value separately for each individual (Girvan & Newman, 2002; Newman, 2006). Individuals differ in their modularity, and the degree of modularity has been found to relate to characteristics such as age (Betzel et al., 2014; Chan, Park, Savalia, Petersen, & Wig, 2014; Chen & Deem, 2015; Geerligs, Renken, Saliasi, Maurits, & Lorist, 2015; Meunier, Achard, Morcom, & Bullmore, 2009a), mental state (Andric & Hasson, 2015; Godwin, Barry, &

Marois, 2015) and neurological status (Alexander-Bloch et al., 2010; 2012; Fornito, Zalesky, & Breakspear, 2015).

Correlations between modularity and performance on a variety of tasks have also been reported, focusing on individuals within a developmental stage. However, there are inconsistencies regarding the direction of the relationship. For example, Stevens et al. (2012) reported a positive correlation between modularity and performance on a visual working memory task, whereas Meunier et al. (2014) reported a negative correlation with odor recognition. Even within the same group of participants, the direction of the correlation varies as a function of task. Alavash et al. (2015) report a positive correlation between modularity and visual working memory performance but a trend towards a negative correlation with a complex working memory span task. Thus, while there is accumulating evidence that whole brain measures relate to behavioral performance (Bassett et al., 2015; Gu, Pasqualetti, et al., 2015a; Gu, Satterthwaite, et al., 2015b; Medaglia, Lynall, & Bassett, 2015; Power et al., 2011; Power & Petersen, 2013), the exact relationship between modularity and performance is unclear.

Deem (2013) proposed a theory derived from theoretical biology that posits high modularity systems afford greater evolutionary fitness at shorter time scales whereas low modularity systems afford greater fitness at longer time scales. This general principle may help to explain why high modularity is beneficial for some tasks while low modularity is beneficial for others. Chen and Deem (2015) developed a computational model to examine this principle in recognition memory. Simulations showed that high modularity architectures resulted in better performance in recognition memory with shorter response deadlines, whereas low modularity architectures resulted in better performance with longer response deadlines. More generally, this principle leads to the prediction high modularity architectures are better for simple tasks, which are completed over a short time scale and draw mainly on a specific module, whereas low modularity architectures are better for complex tasks, which take more time and that draw on the interaction of several modules. Based on this interpretation, the inconsistent previous findings relating modularity to performance might be due to variation in the complexity of the tasks.

Thus, in the present study, we varied task complexity and investigated whether the predicted relation between modularity and complexity on task performance would be observed in a population of young adults. For complex tasks, we selected an operation span task (Unsworth, Heitz, Schrock, & Engle, 2005), a visual arrays short-term memory test (Cowan, 2001; Rouder, Morey, Morey, & Cowan, 2011), an auditory digit span task (Unsworth & Engle, 2007), a cued task-switching paradigm (Kiesel et al., 2010), and a measure of conflict resolution in the domain of visuo-spatial attention from the Attention Network Task (ANT) (Fan, McCandliss, Sommer, Raz, & Posner, 2002). These tasks were identified as complex because each relies on executive attention, that is, the ability to control attention. Operation span and cued task-switching require shifting between tasks. Performance on the visual arrays task has also been demonstrated to relate to attentional control abilities (Shipstead, Lindsey, Marshall, & Engle, 2014). While digit span task is often thought as reflecting passive short-term maintenance, other evidence indicates that it shares many cognitive components with complex span measures, including controlled search of memory, particularly when longer lists are presented for recall (Unsworth & Engle, 2007). The conflict resolution measure taps into the coordination of perception and cognitive control. Simple tasks included two simple visuo-spatial attention measures, alerting and orienting, again from the ANT, and a "traffic light" task in which subjects press a button when a red square turns into a green circle. The alerting and orienting measures from the ANT reflect automatic response to exogenous attentional cues (Corbetta & Shulman, 2002), whereas the traffic light reflects the ability to respond to low level visual properties. Composite simple and complex sores were generated from the behavioral measures by averaging the standard scores with the complex and simple tasks. Given the great differences in the tasks going into the composites, and the variety of abilities that are tapped by each, one would not expect high correlations among the tasks. However, by forming a composite of the measures, one can tap the cognitive mechanisms common to the tasks within a composite and improve the reliability of the dependent measure (Nunnally & Bernstein, 1994; Winer, 1971). These composite measures were correlated with modularity as measured by functional connectivity of resting-state fMRI. According to our theoretical model, the young adults with higher modularity should show better performance on the simple composite, whereas those with lower modularity should show better performance on the complex composite.

## Methods

### Participants

Fifty-two (18-26 years old, Mean: 19.8 years old; 16 males and 36 females) students from Rice University participated in this study. Subjects reported no neurological or psychiatric disorders. Informed consent was obtained according to procedures approved by the Rice University Institutional Review Board. Subjects were compensated $50 for their participation.

### Resting-state fMRI

**Imaging data acquisition.** Resting-state fMRI scans were conducted at the Core for Advanced Magnetic Resonance Imaging (CAMRI) at Baylor College of Medicine. Images were obtained on a 3T Siemens Magnetom Tim Trio scanner equipped with a 12-channel head coil. Foam pads were used to keep subjects' heads stable during the scanning. A high-resolution T1-weighted structural image was acquired by using MPRAGE sequence in the sagittal direction (TR=2500ms,

TE=4.71ms, FoV=256mm, voxel size = 1×1×1 mm$^3$). Three 7-minute resting-state functional runs were obtained by using an echo planar imaging sequence as follows: TR = 2000ms, TE=40 ms, FoV=220 mm, voxel size = 3×3 mm$^2$, slice thickness= 4 mm. Each resting-state run had 210 volumes and each volume had 34 slices in the axial plane to cover the whole brain. All 52 subjects participated in the imaging session.

**Preprocessing.** Image preprocessing was conducted using the AFNI software (AFNI version=AFNI_2011_12_21_1014) (Cox, 1996). The first 6 volumes of each functional run were discarded to allow stabilization of the BOLD signal. Preprocessing procedures basically followed the pipelines recommended by Jo et al.'s paper (Jo et al., 2013) and AFNI's script afni_proc.py. Each functional run was preprocessed separately, including de-spiking of large fluctuations for some time points, slice timing and head motion correction. Then each subject's functional images were aligned to that individual's structural image, warped to the Talairach standard space, and resampled to 3-mm isotropic voxels. Next, the functional images were spatially smoothed with a 4-mm full-width half-maximum Gaussian kernel. A whole brain mask was then generated and applied for all subsequent analysis. A multiple regression model was then applied to each voxel's time series to regress out several nuisance signals, including third-order polynomial baseline trends, six head motion correction parameters, and six derivatives of head motion. [1]

---

[1] Some previous resting-state fMRI studies also included averaged signals in the white matter and ventricle as nuisance signals and regressed them out. A recent study by Aurich et al. (2015) showed that different preprocessing strategies with respect to these tissue-based signals had an impact on some graph theoretical measures, but this study did not evaluate the effect on the modularity measure. Because of Aurich et al.'s findings, we conducted an extra analysis in which the tissue-based (e.g., white matter and ventricle) regression was included in preprocessing by following ANATICOR procedure in AFNI (Jo, Saad, Simmons, Milbury, & Cox, 2010). Specifically, we segmented an individual's structural image by using FreeSurfer (Dale, Fischl, & Sereno, 1999; Fischl, Sereno, & Dale, 1999). Then white-matter and ventricle masks were created and eroded. The averaged time series in eroded ventricle mask and regionally averaged time series within a 15-mm-radius sphere in eroded white matter mask were extracted and included as two additional nuisance signals in regression model. Results in term of modularity-behavior correlations were similar for both analyses without and with tissue regression. In the text we

MODULARITY AND TASK COMPLEXITY   8Bandpass (0.005-0.1 Hz) filtering and outlier censoring were conducted in the same regression model. The outlier censoring removed the time points in which head motion exceeded a distance (Euclidean Norm) of 0.2 mm respect to the previous time point, or in which >10% of whole brain voxels were considered as outliers by AFNI's 3dToutcount. The residual time series after application of the regression model were used for the following network analyses.

**Network Re-construction and modularity calculation.** The whole brain network was re-constructed based on several prominent parcellations of the brain: Brodmann areas, Eickhoff-Zilles atlas (Eickhoff et al., 2005), functional areas identified in Powers et al. (2011), spatially coherent areas identified in Craddock et al. (2012) and a recent published parcellation based on structural and function data from the Human Connectome Project (Glasser et al., 2016). We illustrate the network re-construction methods for the 84 Brodmann areas (42 Brodmann areas for left and right hemispheres respectively). These 84 Brodmann areas were derived from the Brodmann atlas in AFNI Talairach-Daemon (AFNI version=AFNI_2011_12_21_1014) (Lancaster et al., 1997; 2000). This Brodmann atlas mainly covered the cerebral cortex of the brain. Each Brodmann area served as a node of the network, and the mean time series was extracted by averaging the preprocessed time series across all voxels in that Brodmann area. The edge between any two nodes in the network was defined by functional connectivity – that is, the Pearson correlation of the time series for those two nodes. Edges for all pairs of nodes in the network were estimated, resulting in an 84×84 correlation matrix. The correlation matrix was calculated for each run and each subject separately, and an averaged correlation matrix across three runs was obtained for each subject (Figure 1A). We did not concatenate the 3 runs for each subject, because this would introduce a large discontinuity in the fMRI signal at the point of

---

report the results without regressing out tissue-based signals, but the results with this additional preprocessing step are reported in a table available on open science framework.

concatenation, leading to spurious correlation. We applied Newman's algorithm (Newman, 2006) to each individual's averaged correlation matrix to calculate the modularity value for that subject. To do this, we first took the absolute values of each correlation and set all the diagonal elements of the correlation matrix to zero.  Fewer than .05% of the elements in the matrix were negative. The resulting matrix was binarized by setting the largest 400 edges (i.e., 5.67% graph density) in the network to 1 and all others to 0 (Figure 1B). We also considered the top 300 (4.25%) and 500 edges (7.09%), respectively, to assess the consistency of correlation with modularity across different numbers of edges. This binarization process has been argued to improve detection of modularity, by increasing the signal-to-noise ratio (Chen & Deem, 2015). Modularity was defined as in Equation 1 below, where $A_{ij}$ is 1 if there is an edge between Brodmann areas $i$ and $j$ and zero otherwise, the value of $a_i = \Sigma_j A_{ij}$ is the degree of Brodmann area $i$, and e = ½ $\Sigma_i a_i$ is the total number of edges, here set to 300, 400, and 500 respectively. Newman's algorthim was applied to the binarized matrix to obtain the (maximal) modularity value and the corresponding partitioning of Brodmann areas into different modules for each subject. [2]

$$M = \frac{1}{2e} \sum_{all\ module\ areas\ i,j\ within\ this\ module} \left(A_{ij} - \frac{a_i a_j}{2e}\right) \qquad (1)$$

A group-average correlation matrix (Figure 1A) was obtained by averaging the mean correlation matrices across all subjects. The partitioning into modules was also conducted at the group level. To assess individual variability in the modular structure of the brain, we maximally aligned each subject's partitioning of Brodmann areas with the partitioning identified from the group-average correlation matrix. When comparing the distance between two assignments of Brodmann areas to modules, we take into account that the labeling of the modules can be permuted and that the number of modules of the two assignments need not be identical. Thus the maximal alignment is the one that minimizes the distance over all permutations of the assignment with the greater (or equal) number of modules. The value $d$ is the number of nodes whose modular assignment differed between the subject and the group average, defined by Equation 2.

---

[2] Due to the resolution limitation issue with the current community detection algorithm (Fortunato & Barthélemy, 2007), we also estimated modularity values by using the lowest hierarchy level in the Louvain algorithm (Blondel, Guillaume, Lambiotte, & Lefebvre, 2008), which could avoid resolution limit issue (Lancichinetti & Fortunato, 2009; 2014). Results in term of modularity-behavior correlations based on Louvain method are similar to those with modularity calculated by the method described above and are reported in a table that is available at the open science framework.

$$d = \min_{P} \sum_{i=1}^{84} (1 - \delta_{m1_i, Pm2_i}) \qquad (2)$$

Brodmann areas are indexed by $i$, $m1_i$ denote the module to which Brodmann area $i$ belongs in assignment 1, $m2_i$ denote the module to which Brodmann area $i$ belongs in assignment 2, $P$ is a permutation of module labeling, and we have assumed the number of modules in assignment 1 is not greater than that in assignment 2. To assess which cortical regions drove this individual variation in modular structure, we also calculated, for each of the 84 Brodmann areas, the number of subjects for whom that Brodmann area was assigned to a different module than the assignment for group-average data.

While Brodmann areas have the virtue of covering most of the gray matter, many of the areas are quite large, meaning that each Brodmann area likely contains multiple functionally selective regions. Given that each Brodmann area can only be assigned to one module, that modular assignment may only be appropriate for a subregion, thereby potentially reducing the relation between modularity and behavioral performance. Thus, we also calculated modularity for each subject based on another anatomical parcellation which is available in AFNI (Eickhoff-Zilles macro labels atlas from N27 template in Talairach space) (Eickhoff et al., 2005), and some functional parcellations, including Power et al.'s (2011), Craddock et al.'s (2012) and Glasser et

al.'s (2016) networks. We used 90 of the 115 masks from the Eickhoff-Zilles atlas, excluding those in the cerebellum. Power et al. defined many small spheric nodes that are associated with specific cognitive functions identified by meta-analyses of task-based fMRI [3]. Craddock et al. parceled the gray matter into many spatially coherent clusters, identifying clusters based on homogeneity of local functional connectivity. We chose the 100-node version of Craddock et al.'s parcellation, which has a similar number of nodes with the Brodmann network. Glasser et al. used multiple properties (e.g., cortical myelin content, cortical response to task functional MRI, resting-state functional connectivity pattern) of MRI data from Human Connectome Project and generated feature gradient maps to identify the boundaries of structurally and functionally homogenous areas on the cortical surface model, and then created 180 areas in each hemisphere [4]. The correlation matrices for functional parcellations were binarized with a threshold of percent cost (5.67%), which equates the proportion of top edges to that for the Brodmann network of 400 edges.

**Behavioral tasks**

All fifty-two subjects participated in the operation span and task-shifting tasks. Forty-three of them participated in the ANT task and visual arrays task, and forty-four subjects participated in the traffic light task, as these were done in a different session, and not all subjects returned to participate in all tasks. The interval between neuroimaging and behavioral sessions varied from 0 (i.e., measuring resting-state fMRI and behavior on the same day but during different sessions) to 140 days. 40 subjects completed all of the behavioral tasks.

---

[3] The original Power et al.'s network consisted of 264 nodes. To apply it to our data, we firstly converted the coordinates provided in Power et al.'s paper from MNI space to Talairach space by using a converter tool (Lancaster et al., 2007) in software GingerALE. Then, 10mm-diameter spheric ROIs were created around coordinates. Eight ROIs in the cerebellum in Talairach space were excluded as they were not covered by images for some of our subjects, resulting 256 nodes in our analyses.

[4] The original Glasser et al.'s parcellation was based on the two-dimensional cortical surface model. We obtained the volumetric parcellation (https://figshare.com/articles/HCP-MMP1_0_projected_on_MNI2009a_GM_volumetric_in_NIfTI_format/3501911) which was projected to the MNI space. To apply this volumetric parcellation to our data, using AFNI script @auto_tlrc, we firstly warped the anatomical template onto which this functional parcellation was co-registered into Talairach space and registered it to the template used in AFNI with linear interpolation. Then a set of masks for volumetric parcellation were warped into Talairach by using the same transformation, and resampled to match the resolution of our data with nearest neighbor interpolation. Craddock et al.'s parcellation was applied to our data in a similar way.

**Operation Span.** Subjects were administered the operation span task (Unsworth et al., 2005) which is a standard measure of working memory capacity with high test-retest reliability (Redick et al., 2012). On each trial, participants saw an arithmetic problem for 2 seconds, e.g., (2×3)+1, and were instructed to solve it as quickly and accurately as possible. Subjects judged whether a digit presented on the next screen was a correct solution to the previous arithmetic problem. After the arithmetic problem, a letter was presented on the screen for 800ms that subjects were instructed to remember. A single trial consisted of repeating these steps six or seven times. At the end of each trial, subjects recalled the letters in the order in which they were presented. The recall screen consisted of a 3×4 matrix of letters and subjects used the mouse to check the boxes aside letters during recall. There were twelve trials for each set size (six or seven arithmetic problem – letter pairs, randomly presented), resulting a total of 156 letters and 156 arithmetic problems. The dependent measure was the total number of letters recalled at the correct position with a maximum of 156. Prior to the experiment, participants practiced with (1) a block involving only letter recall, (2) a block involving only arithmetic problems, and (3) a mixed block in which the trial had the same procedure as in the experimental trials. Practice blocks included only set sizes of 2, 3, or 4.

**Visual Arrays Task.** A visual array task was used to tap visual short-term memory capacity. In this task, subjects fixated the center of the screen. Arrays of 2 to 5 colored squares at different screen locations were presented for 500ms, followed by a blank screen for 500ms, and then by multi-colored masks for 500ms. A single probe square was then presented at one of the locations where the colored squares had appeared. Subjects had to judge whether the probe square had the same color as the one at the same location. The order of different array sizes was random. There were 32 trials at each array size, half of which required positive responses and half negative. The visual short-term memory score was calculated by averaging the accuracy across all array sizes.

**Digit Span Task.**  Subjects heard a list of spoken digits, at the rate of one per second. After the last digit in each list, a blank screen prompted participants to recall the digits in the order presented by typing them on the keyboard. Subjects were tested on five trials for each set size starting at set size two to a maximum of nine. The program terminated if participants scored fewer than 3 correct trials at a given set size (60% accuracy). Digit span was calculated by estimating the set size at which a subject would score 60% through linear interpolation between the two sizes spanning this accuracy threshold.

**Task-Shifting Task.**  Participants responded to an object (blue and yellow squares and triangles) according to a preceding cue word. If the cue was "color", participants pressed a button to indicate whether the object was blue or yellow, and if the cue was "shape", they pressed a button to indicate whether the object was a square or a triangle. The same buttons were used for the two tasks. Response time was recorded from the onset of the object. For half of the trials, the cue was the same as that in the previous trial (a repeat trial) and for the other half, the cue changed (a switch trial). For each condition, to take into account both response time and accuracy in a single measure, we calculated an inverse efficiency (IE) score (Townsend & Ashby, 1983) defined as mean RT/proportion correct. The task shifting cost was measured as the difference in inverse efficiency score between repeat and switch trials. Cue-stimulus Interval (CSI) was varied, which is the time between onset of the cue and onset of the object, using CSIs of 200, 400, 600, and 800 ms. However, as the effect of modularity on IE did not differ for different CSIs, the data were averaged across CSI. In total, there were 256 repeat trials and 256 switch trials.

**Attention Network Test.** The Attention Network Test (ANT) (Fan et al., 2002) was used to measure three different attentional components: alerting, orienting, and conflict resolution. This task involves a simple visual discrimination– that is, indicating whether a central arrow is pointing left or right via a mouse press. The arrow(s) appeared above or below the fixation cross, which was in the center of the screen. The central arrow appeared alone on a third of the trials and was flanked by two arrows on the left and two on the right on the remaining two third of trials. The flanking arrow trials were evenly split between a congruent condition in which they pointed in the same direction as the central arrow and an incongruent condition in which they pointed in the opposite direction. In the neutral condition, there were no flanking arrows. The timing of the appearance of the arrow(s) was cued by zero, one or two asterisks, which appeared for 100ms on the screen. The interval between offset of the cue and onset of the arrow was 400ms. The four cue conditions were: 1) no cue condition, 2), a cue at fixation, 3) double-cue condition, with one cue above and the other below fixation, 4) spatial-cue condition, where the cue appeared above or below the fixation to indicate where the arrows would appear, provided both timing and location information. Thus, the task had a 4 cue $\times$ 3 flanker condition factorial design. The experimental trials consisted of three sessions, with 96 trials in each session, and 8 trials in each cell of each condition. For half of all trials, arrows were presented above the fixation and for the other half below. Also, for half of the trials, the middle arrow pointed left and for the other half right. The order of trials in each session was random. Before the experimental trials, 24 practice trials with feedback were given to subjects.

Response times and accuracy were recorded with mean RT based on correct trials only. Inverse efficiency (IE) was calculated for each condition. The alerting effect was computed by subtracting the IE for the no cue condition from the IE for the double cue condition. The orienting effect was computed by subtracting the IE for the center cue condition from the IE for the spatial cue condition. The conflict effect was computed by subtracting the IE for the congruent condition from the IE for the incongruent condition. To make the direction of the conflict effect the same as alerting and orienting effects, we reversed the sign of the conflict effect. Thus, the more negative the conflict effect value, the greater the interference from the incongruent flankers. The alerting and orienting effects both tap into sensitivity to exogenous, or stimulus-driven, cues within the visual domain, while the conflict effect taps individuals' ability to direct attention to the central cue and ignore conflicting flankers, thus drawing on endogenous, or subject-driven, attentional abilities (Raz & Buhle, 2006).

**Traffic light task.** In this task, subjects saw a red square in the center of screen, which was replaced after an unpredictable time delay (from 2 to 3 seconds) by a green circle. Subjects pressed a button as quickly as possible when they saw the green circle. There were 25 trials in total. Mean response time was calculated for each subject.

**Composite scores and correlation analyses**

Split-half reliability for each task was calculated using the Spearman-Brown prophecy formula (Allen & Yen, 1979). As discussed in the results, the reliabilities were moderate to high for all except for that for the alerting component of the ANT. The low reliability for alerting is consistent with previous studies showing higher reliabilities for the orienting and conflict components of the ANT than for alerting (Fan et al., 2002). Given the absence of reliability for this measure, it was eliminated from consideration. Two composite scores were calculated: one for complex tasks and another for simple tasks. In order that higher numbers would be associated with better performance for all tasks, the values for shift cost and traffic light were reverse scored by multiplying by -1. Then, for each task, standard scores (z-scores) were obtained based on the 40 subjects who had scores on all eight tasks. A complex composite score was calculated by averaging the z-scores across the five complex tasks: operation span, visual array task, digit span, task-shifting task, and the conflict measure of the ANT task. A simple composite score was calculated by averaging the z-scores across two simple tasks: orienting from the ANT and the traffic light task (RT). Participants' modularity values were correlated with their complex and simple composite scores. Based on the predictions of the theoretical model, we expected that modularity would correlate positively with the simple composite score, and negatively with the complex composite score.

## Results

**Modular organization of resting-state network**

For the calculations with the BA-network 400 edges (5.67%), modularity values ranged across subjects from 0.33 to 0.59, with a mean of 0.47 on a scale from 0 to 1.0. For the BA-network 300 (4.25%) and the BA-network 500 edges (7.09%) the means were 0.53 (range: 0.39-0.64) and 0.42 (range: 0.31-0.52), respectively. With the BA-network 84 nodes and 400 edges used in our analysis, a value greater than 0.29 is taken to indicate a significant degree of modularity as random voxel data led to mean $M = 0.25$ (SD = 0.022, max value out of 52 = 0.29) for our setup. The results reported below were based on the BA-network 400 edges (5.67%) without tissue-based regression, unless indicated otherwise. Individuals differed to some degree in the number of modules, with most (31/52) having four modules but with some having three modules (14/52) or five modules (7/52). The organization of brain regions into modules is described below, starting with the results from the group average data, followed by a discussion of how this organization varied across individuals.

At the group level, the 84 Brodmann areas were organized into four modules (Figure 1C, D). For all modules, the brain regions making up the module were contiguous and symmetrical, with the left and the right Brodmann areas assigned to the same module for forty of the forty-two BA labels. The first module (a sensory-motor module) consisted of Brodmann areas covering bilateral primary somatosensory cortex (BA 1, 2, &3), primary motor cortex (BA4), somatosensory association cortex (BA 5), premotor cortex and supplementary motor cortex (BA 6), supramarginal gyrus (BA 40) and ventromedial prefrontal cortex (BA 25) [5]. The second module included auditory and language-related regions covering both Broca's and Wernicke areas, though areas were bilateral, consisting of bilateral auditory cortex (BA 41, 42), superior temporal gyrus (BA 22), temporal pole area (BA 38), insular gyrus (BA 13), primary gustatory cortex (BA 43), ventral part of dorsolateral prefrontal cortex (BA 46), and inferior frontal gyrus, including pars opercularis (BA 44), pars triangularis (BA 45), and pars orbitalis (BA 47). The third module overlapped considerably with the default mode network (Raichle et al., 2001), consisting of bilateral angular gyrus (BA 39), inferior temporal gyrus (BA 20), middle temporal gyrus (BA 21), some frontal regions, including frontal eye fields (BA 8), the dorsal part of dorsolateral prefrontal cortex (BA 9), anterior prefrontal cortex (BA 10), as well as some medial regions, such as anterior cingulate gyrus (BA 24, 32, & left BA 33) and a part of posterior cingulate gyrus (left BA 31). The fourth module was the largest and most diverse. It included visual, attentional and medial memory-related areas. More specifically, this module included bilateral orbitofrontal cortex (BA 11), posterior superior parietal cortex (BA 7), primary, secondary and associative visual cortex (BA 17, 18, & 19), fusiform gyrus (BA 37), as well as some medial regions, including the posterior cingulate cortex (BA 23, 29, 30, right BA 31), a part of anterior cingulate gyrus (right BA 33), piriform cortex (BA 27), entorhinal cortex (BA 28, 34), and perirhinal cortex (BA 35, 36). The four modules roughly replicated prior results on various parcellation schemes (Brodmann areas using Newman's algorithm: Chen & Deem, 2015; AAL atlas using Newman's algorithm: He et al., 2009; $8\times8\times8$ mm$^3$ cubic nodes within the coverage of AAL atlas using Louvain algorithm: Meunier et al., 2009b; 2010). We also obtained group-level modular organizations for our data for other parcellations, which revealed four modules for Power's and Craddock's nodes, three for Glasser's nodes, and five for the Eickhoff-Zilles nodes. The modular organization was largely consistent across these different networks for three of the four modules from Brodmann areas (i.e., the sensory-motor module, the default-mode module, and the diverse module which covered the occipital, dorsal parietal and some frontal regions); however, the auditory/language module was not evident in all other networks (for modular organizations in other networks, see Supplementary Figure 2 available at the open science framework: https://osf.io/9tmhc/?view_only=7bafc4494c8e49f8a1b8c51df5823b5a). Hence, the number of modules and their constituent brain regions were quite similar irrespective of whether an anatomical or functional parcellation scheme was used, though some differences were observed.

---

[5] Left and right BA 25 had no connection with any other nodes when we binarized the correlation matrix, resulting random assignment of these two nodes into this module.

While the above modules represent those derived from the average correlation matrix, the assignment of brain regions to modules differed considerably across subjects, with some brain regions showing a high degree of variation and others showing little. Figure 2A shows for each brain area the number of individuals for which that area was in a module different from the average. In low-level sensory and motor area, e.g., posterior visual cortex and primary somatosensory and motor cortex, the assignment to a module differed from the group average for only a few participants. In contrast, for brain regions associated with higher order cognitive functions, e.g., lateral prefrontal cortex, there was large variation across subjects. In some temporal and parietal regions, module identity differed from the group average for about half of the subjects. The medial ventral frontal lobe had the most subjects deviate from the group average, which most likely was due to the distorted magnetic field in this region, resulting in a poor signal-to-noise ratio with the typical EPI sequence. Notably, however, cognitive control regions that do not experience such signal loss also showed considerable variability in module assignment across individuals.

To quantify individuals' difference in modular organization relative to the group average, we computed the average distance of that individual's modular organization to the modular organization of the group-average data, as described above. Distance ranged from 14 to 48 with a mean of 33.8. Figure 2B shows the relation between individual's modularity value ($M$) and this distance measure and includes a depiction of the distribution of both measures. There was a significant negative correlation ($r = -0.36$, $p = 0.008$) between an individual's modularity score and their distance from the group-average modular structure, indicating that the modules identified for high modularity individuals were closer to the group-average modules, than were those for low modularity individuals.

**Behavioral and correlation analyses**

For the operation span task, one subject was excluded from the analysis due to a very low score (27 out of a maximum of 156), which was more than three standard deviations below the group mean (mean = 123, SD: 24). For the task-shifting task, two subjects were excluded as one performed at a chance level of accuracy and the other showed a shift costs more than three standard deviations greater than the group mean. Overall, there was a significant task shifting effect (mean shift cost = 194, SD: 94; $t(49) = 14.54$, $p < 0.0001$). Forty-three subjects participated in the visual arrays task and the ANT task. The mean accuracy for visual arrays was 89.5%, with a standard deviation of 5.4%, which is similar to that from prior reports (Cowan, 2001).  Forty-four subjects participated in the digit span task. The mean capacity for digit span was 7.26, with a standard deviation of 0.94. For the ANT task, one subject's data for the conflict task were excluded as the conflict effect was more than four standard deviations greater than the group mean. Overall for the ANT task, the mean findings replicated prior results (Fan et al., 2002) as there was a significant alerting effect (mean = 43, SD: 28; $t(42) = 10.14$, $p < 0.001$), a significant orienting effect (mean = 49, SD: 29; $t(42) = 10.89$, $p < 0.001$), and a significant conflict effect (mean = 139, SD: 42; $t(41) = 21.26$, $p < 0.001$). Forty-four subjects participated in the traffic light task. The group mean RT was 227ms, with a standard deviation of 17ms. All behavioral tasks except for alerting had medium to high reliabilities.  Specifically, the Spearman-Brown prophecy reliabilities were 0.85 for operation span, 0.84 for the visual arrays task, 0.90 for digit span, 0.68 for shifting task, 0.55 for conflict measure of ANT task, 0.38 for orienting measure of ANT task, and 0.94 for the traffic light task.   However, the reliability for the alerting measures from the ANT task was -0.31.  Thus, this measure was eliminated from further consideration.

     Composite scores were generated based on 40 subjects who had scores on all tasks. We examined the pattern of correlations between the task scores and the composites and between the two composites to verify that these composites tapped different underlying cognitive constructs. Specifically, we found that all five complex tasks' measures correlated significantly with the complex composite score ($r = 0.63$, $p < 0.001$ for operation span; $r = 0.49$, $p = 0.001$ for visual

arrays; $r = 0.72$, $p < 0.001$ for digit span; $r = 0.58$, $p < 0.001$ for conflict from the ANT task; $r = 0.37$, $p = 0.018$ for shifting), but were not significantly correlated with the simple composite score ($r = -0.25$, $p = 0.13$ for operation span; $r = 0.27$, $p = 0.09$ for visual arrays; $r = 0.08$, $p = 0.63$ for digit span; $r = 0.06$, $p = 0.7$ for conflict from the ANT task; $r = -0.15$, $p = 0.34$ for shifting).  For the simple composite score, given that only two measures went into the composite, the correlations between the individual measures and the composite were necessarily high and equivalent ($r = 0.73$, $p < 0.001$ for orienting measures of the ANT task; $r = 0.73$, $p < 0.001$ for the traffic light task),   The scores from the simple tasks were not significantly correlated with the complex composite score ($r = 0.03$, $p = 0.84$ for orienting measures of the ANT task; $r = -0.03$, $p = 0.87$ for the traffic light task). The correlation between the simple and complex composite scores was near zero ($r = 0.005$, $p = 0.98$).

Figure 3 depicts the relation between modularity and the composite scores, with the complex composite score on the right and the simple composite score on the left. The results for both conformed to predictions, a marginally significant negative correlation between complex composite scores and modularity ($r = -0.29$, $p = 0.067$) and a positive correlation between the simple composite scores and modularity ($r = 0.41$, $p = 0.008$). (Correlations between modularity with each of the task measures are available online at the open science framework: https://osf.io/9tmhc/?view_only=7bafc4494c8e49f8a1b8c51df5823b5a). A test for the difference in dependent correlations (Meng, Rosenthal, & Rubin, 1992) showed that the two correlations differed significantly ($z = 3.09$, $p = 0.002$), confirming the relation between modularity and performance was mediated by task complexity. We also correlated the complex and simple composite scores with the modularity values derived from the binarized matrices based on 300 and 500 edges, and similar patterns were observed: 1) 300 edges (4.25%): $r = -0.32$, $p = 0.044$ for complex score and $r = 0.33$, $p = 0.037$ for simple score, difference: $z = 2.84$, $p = 0.0045$, 2) 500 edges (7.09%): $r = -0.30$, $p = 0.059$ for complex score and $r = 0.34$, $p = 0.032$ for simple score, difference: $z = 2.80$, $p = 0.0052$. Thus, the pattern of correlations was highly consistent across differences in graph density. [6]

    Modularity derived from another anatomical network (Eickhoff-Zilles) had a marginally significant negative correlation with the complex composite score ($r = -0.305$, $p = 0.056$) and a non-significant positive correlation with the simple composite score ($r = 0.19$, $p = 0.24$). The difference between two correlations was significant ($z = 2.16$, $p = 0.03$). This pattern basically replicated that from the Brodmann network. Modularity derived from Power et al.'s network (Power et al., 2011) had a non-significant negative correlation with the complex composite score ($r = -0.25$, $p = 0.12$) and a near zero correlation with the simple composite score ($r = -0.057$, $p = 0.73$). The difference between the two correlations was not significant ($z = 0.85$, $p = 0.40$). These findings most likely result due to the much lower coverage of gray matter for the Power's nodes than for the Brodmann areas. Based on our calculations of the total number of voxels of the brain areas covered by the masks from the two set of nodes, the Brodmann areas covered about three times as much gray matter as the areas defined by the Power's et al. (2011) nodes. For modularity based on Craddock et al.'s 100-node parcellation and a recently released functional parcellation with greater gray matter coverage (Glasser et al., 2016), patterns of correlations with the behavioral composite scores were similar to that for the Brodmann network, but the correlations did not reach significance ($r = -0.217$, $p = 0.18$ for the complex composite and $r = 0.154$, $p = 0.343$ for the simple composite for Craddock's network; $r = -0.257$, $p = 0.109$ for the complex composite and $r = 0.151$, $p = 0.352$ for the simple composite for Glasser's network).

---

[6] The results for the tissue-based regression and using the Louvain algorithm for computing modularity also gave similar results, which are reported in a table that is available at the open science framework.

However, there was a marginally significant difference between the two correlations ($z = 1.77$, $p = 0.077$) for Glasser's network, but the difference did not reach significance for Craddock's network ($z = 1.61$, $p = 0.108$). Aside from the Power's et al. network, the correlations with behavior for different parcellation schemes were similar in showing a negative correlation with the complex composite and a positive correlation with the simple composite, with the difference between the two being significant or marginally so.

Given that the interval between resting-state fMRI session and behavioral testing session varied substantially across participants, we investigated whether the strength of correlations with modularity differed as a function of this interval. For each task, we regressed the dependent measure on modularity, absolute value of the number of days between resting-state fMRI and task completion, since for a few subjects tasks were completed before fMRI data were collected, and the interaction of these two variables. Rather than perform this analysis with all of the dependent measures of modularity described above, we used only the BA-network 400 edge modularity measure, as this network showed the strongest correlations with the simple and complex composite scores when we collapsed time interval.

Among the complex tasks, operation span and conflict from the ANT task showed a decreasing correlation with modularity as the number of days increased, with the interaction between modularity and days being significant for operation span ($p=0.02$) and marginally so for conflict ($p=0.07$). For both tasks, correlations were moderately high at the shortest interval but declined substantially over days. Task-shifting did not show a significant interaction between modularity and days ($p = 0.22$). However, given that operation span and shift cost had the largest range of days and included the shortest intervals, we also computed a composite score combining standardized operation span and shift costs scores into a single measure. For this composite measure, there was a significant interaction between days and modularity ($p=0.04$), with correlations within days groups ranging from -0.59 ($p=0.04$) for day group 1 to 0.04 ($p=0.89$) for day group 4.

For the remaining tasks, visual STM and digit span showed no sign of an interaction between modularity and number of days (visual STM: $p = 0.89$; digit span: $p = 0.92$) nor did any of the simple tasks (orienting and traffic light: both $p$s $> 0.73$). It is unclear whether this failure to find an interaction relates to specifics of these tasks or to the fact that all of the day intervals, including the shortest one, were quite long.

Figure 4 depicts relations between modularity and the dependent variable as a function of days. Subjects were divided into day groups such that the number of subjects in each group was approximately equal, and then the correlations between the dependent measure and modularity were plotted for each day group. For operation span and task shifting, there were four such day groups (group 1: 0-11 days, $N=14$; group 2: 11-31 days, $N=12$; group 3: 33-43 days, $N=12$; group 4: 44-82 days, $N=14$). For the ANT task, visual arrays, traffic light, and digit span, there were three day groups (group 1: 30-43 days, $N=13$; group 2: 44-82 days, $N=14$; group 3: 100-140 days, $N=16$). Note that groups 1 and 2 for the latter tasks have approximately the same range of days as for groups 3 and 4 for operation span and shifting. The figure includes the three complex tasks that showed significant or marginally significant interactions between days and modularity and also the results for the orienting effect, which showed a significant positive correlation with modularity overall with no change across day groups.

## Discussion

The current study examined the application to brain organization of a theory derived from theoretical biology, which posits that high modularity systems are advantageous over short time scales, whereas low modularity systems are advantageous over longer time scales. In testing this theory with respect to the relation of brain organization to behavior, we related time scale to task complexity, as simpler tasks have fewer cognitive steps and therefore take less time to complete than do more complex tasks that entail a greater number of steps. Using Newman's algorithm to segregate brain regions into modules, we determined the modularity value for each subject and correlated modularity with behavioral performance, testing whether we would find the predicted interaction between complexity and modularity. The modules uncovered in this process consisted of subnetworks of contiguous regions that had fairly transparent mappings to cognitive functions. Considerable variability in terms of the identity of modules and the overall modularity value was found across subjects. The findings for task correlations were generally consistent with the proposed theory. Below, we summarize the findings on modularity and relation of modularity to behavior and discuss their implications for theory.

**Modular brain network and individual differences**

The four modules uncovered at the group level matched to a large extent those uncovered in prior studies using modularity analyses of resting state and task-based fMRI (Chen & Deem, 2015; He et al., 2009; Meunier et al., 2009b; 2010). It is also the case that the modules detected using graph theoretical tools agree with modules discovered by other analytical approaches, such as independent component analysis (Smith et al., 2009) or clustering analysis (Blank, Kanwisher, & Fedorenko, 2014; Yeo et al., 2011). Taken together, these findings indicate that the human brain is intrinsically organized into functionally different modules.

While the group data revealed a coherent set of modules, there was considerable variability across individuals in terms of the assignment of brain regions to modules, with interesting patterns regarding which regions were more or less likely to vary in module assignment. Specifically, brain regions involving motor or primary or secondary sensory processes, such as visual, somatosensory, or auditory, showed little variability. On the other hand, regions associated with higher-level functions like cognitive control or working memory, i.e., dorsolateral prefrontal or parietal regions, showed much more variability. These findings are consistent with claims that these high-level regions are domain-general, and can be associated with a variety of sensory or motor processes or representations depending to the current demands of the system. It is unclear, however, why these domain-general regions might be preferentially associated during resting-state with different sensory or motor regions for different individuals, for instance, auditory regions for one individual but visual regions for another. Perhaps these alliances depend on whatever individuals tended to think about during their resting-state scans or could perhaps reflect more permanent preferences in allocation of attention due to greater experience or interest in certain domains over others. There was also a tendency for individuals with lower modularity to deviate more from the group-average modules than individuals with higher modularity (Figure 2B). Less modular individuals have, by definition, a smaller difference between the within-module correlations and the between-module correlations. On explanation for this trend, therefore is that for less modular individuals it is more likely that modules are more loosely defined or that Newman's algorithm will identify an idiosyncratic set of nodes to maximize the difference between within-module and between-module correlations.

**Modularity of brain network and behavioral performance**

As depicted in Figure 3, a strikingly different pattern of variation in the direction of correlations for complex versus simple tasks was observed, which conformed with predictions of a positive correlation with the composite score for the simple tasks and a negative correlation with the composite score for complex tasks. This crossover interaction between task-complexity and brain modularity can be interpreted with respect to how tasks make use of a combination of domain-specific and domain-general processes. Previous task-related functional imaging studies have found that not only domain-specific regions, such as those for speech or visuo-spatial attention (Martin, 2003; Noudoost, Chang, Steinmetz, & Moore, 2010; Price, 2010; 2012; Squire, Noudoost, Schafer, & Moore, 2013), but also domain-general attentional or cognitive control regions are activated during the performance of complex tasks involving, for instance, language comprehension or working memory (Fedorenko, Duncan, & Kanwisher, 2012; 2013; Nee et al., 2013). Moreover, the same patterns of synchronization of those regions have been reported even when there is no specific task (i.e., resting-state fMRI) (Fox & Raichle, 2007; Fox, Corbetta, Snyder, Vincent, & Raichle, 2006; Xiang, Fonteijn, Norris, & Hagoort, 2010). These latter findings indicate that such co-activations or synchronization do not arise solely from the demands of specific experimental paradigms, but reflect the intrinsic connectivity between regions in a network, and further suggests that such interactions among domain-specific and domain-general modules of brain organization underlie cognition. Based on this view, both domain-specific or domain-general functions are supported by the within-module connections, and interactions between domain-specific and domain general regions are supported by between-module connections. Thus, it is reasonable to assume that any behavioral task relies on both within-module and between-modules connections, but with different relative weights for each depending on task complexity. A simple task may rely more on the within-module but less on between-modules connections, whereas a complex task may rely more on the between-modules but less on the within-module connections. In fact, the measure of modularity reflects these relative weights of within-module and between-modules connections.  Thus, the positive correlation between the simple composite scores and modularity suggests that the strong connections within a module relative to between modules favor low-level processing that draw on that module. Conversely the negative correlation between the complex composite scores and modularity suggests that relatively weak connections within a module but stronger connections between modules promote high-level processing, e.g., some high-level regions which govern or flexibly control low-level regions, e.g., allowing switching between arithmetic or encoding information into memory in the operation span task.

**Time-sensitive correlations**

One unanticipated finding from the current study was that the relationship between modularity and task performance depended on the time interval between neuroimaging and behavioral testing. This result is clearest with the complex tasks of operation span, task shifting, and the conflict measure from the ANT. In all three cases, the strongest correlations were observed for short intervals, with correlations approaching zero at long intervals.

One possible explanation for these time-sensitive correlation patterns is that neither modularity nor task performance are stable individual differences. Rather, these measures may all drift over time. Poldrack et al. reported substantial fluctuations in the connectome structure of a single individual over the course of 18 months of testing (Poldrack et al., 2015). The stability of some of the behavioral measures, for example operation span, has also been previously investigated. While Klein and Fiss (1999) reported significant test-retest correlations on operation span with, approximately 3 week, 6-7 week and 10 week intervals, the raw test-retest correlations were lowest with the longest interval. At a given time point, both brain and behavior measures are reliable, which can explain why a correlation is observed between modularity and task performance with a short time interval. But as time passes, the modular organization of the brain changes, as does task performance, in a sort of random walk that drifts over time. Thus, after a long interval, the correlation between modularity and task performance weakens. To test this hypothesis, future research should investigate the stability of both resting-state modularity and behavioral measures, as well as the stability of the brain-behavior correlations over different intervals. Our results suggest that rather than measures of an individual's network structure being a fingerprint – a stable, identifying property of an individual (Finn et al., 2015) – they are probably more like a snapshot, a reflection of how the brain was organized at a specific time. Understanding how and why brain organization changes over time remains an important area for research.

**Limitations and future directions**

While the pattern of results across different parcellation schemes was similar (aside from the Power et al., 2011, network), the strength of the correlations with the complex and simple composites and the significance of the difference in correlations varied.  The strongest results were obtained for the two anatomical parcellation schemes, with both revealing a significant difference in the correlations with the simple and complex composites in the direction predicted. It is unclear why the results for the parcellations taking into account functional connectivity resulted in weaker results. Since previous work argued that inappropriate node definition might mischaracterize brain regions which have distinctive functions (Wig, Schlaggar, & Petersen, 2011), and found that nodes derived based on task-based fMRI studies did not align well with anatomical parcellations (Power et al., 2011), one might have expected the opposite – that is, that using network nodes determined at least in part from functional activations (e.g., Glasser et al., 2016) would result in modular structures more relevant to task performance and thus higher correlations of modularity with behavior. However, while the correlations from the functional parcellations were often not substantially lower than those determined from the Brodmann parcellation, they were never substantially higher. It should be noted that the only other study correlating behavior on simple and complex tasks to graph theoretic measures (Cohen & D'Esposito, 2016) also did not find consistently higher correlations using a functional than an anatomical atlas (e.g., for their relatively simple sequence tapping task, the correlation with modularity from an anatomical parcellation was significant, $p = 0.03$, whereas that for a functional parcellation was not, $p = 0.40$).   One factor to keep in mind is that modularity determined from Newman's algorithm does not use modules determined on a priori functional grounds when calculating modularity, irrespective of the network used. Instead, the connectivity determined from resting state activity is used to determine how nodes should best be sorted into modules in order to maximize modularity. Across the different parcellation schemes examined in our study, the number of modules and their constituent brain regions were quite similar for the most part.  It must be left to future work to determine exactly how the differences that did exist in modular structure arose and what the implications of these differences were for the determination modularity values.  Another issue that needs to be addressed is whether the combination of quite different tasks into the complex and simple composites negated any advantage for network nodes determined from particular functional domains.  In future work, it will be important to manipulate task complexity within a given domain (e.g., manipulating working memory demands parametrically).  It would also be important to determine if using a set of nodes limited to those domain-specific (e.g., auditory processing) and domain-general (e.g., exogenous attention) networks hypothesized to be involved in a particular task would result in higher correlations with behavior than whole brain measures. Perhaps under these conditions, modularity determined from subnetworks defined by functional parcellation schemes would result in higher correlations with behavior than those defined anatomically.  On the other hand, it is possible that whole brain measures would remain superior because connectivity to other regions plays some understudied role – for instance, in the suppression of unrelated domain-specific or domain-general subnetworks (Todd, Fougnie, & Marois, 2005).

## Conclusion

The relationship between individual differences in whole brain modularity and performance varies as a function of task complexity. Specifically, low modularity individuals showed an advantage on complex tasks while high modularity individuals showed an advantage on simple tasks. These observations support predictions of a theoretical model, by which high modularity increases performance for simple tasks that operate at a short time scale, and low modularity increases performance for complex tasks that operate at a longer time scale. This new understanding allows previously reported, and conflicting, results relating individual differences in brain organization to behavior to be understood in a general framework. Future work is needed in order to examine whether modularity as defined by subnetworks of functionally relevant systems would be superior to these whole brain measures or rather than the whole brain measures would maintain an advantage due to the involvement of other regions whose contributions is less clear based on current findings.

**Acknowledgements**

This work was partially supported by the T.L.L. Temple Foundation. MWD and FY were partially supported by the Center for Theoretical Biological Physics under NSF grant #PHY-1427654. We would like to acknowledge the contributions Jill Moore, Yuan Chang and Elizabeth Baca made to data acquisition and analysis.

# Figure Captions

**Figure 1**. (A) The group-average correlation matrix in which each element represents the edge (i.e., functional connectivity) between each pair of 84 Brodmann nodes. (B) The group-average binarized correlation matrix by setting the off-diagonal top 400 edges of correlation matrix in (A) to 1 whereas all others to 0. (C) The group-average whole brain modular organization with the Brodmann areas in each module being marked by the same color. The modular partition results were mapped onto a brain surface model for display by using SUMA in AFNI. (D) The group-average binarized correlation matrix sorted by the group-level modular partition. The modular color markers are the same as in (C).

**Figure 2**. (A) The distance map showing how many subjects' module identities differ from group-average in each Brodmann area. (B) The scatterplot showing average distance across 84 Brodmann areas for each subject against modularity value, as well as the distributions for two measures.

**Figure 3**. The relation between modularity and the composite scores with the simple on the left and the complex on the right. The center of the figure depicts the theoretical prediction from Chen and Deem (2015) relating performance to tasks at different levels of complexity for individuals with high and low modularity. The blue lines represent the linear fit.

**Figure 4**. Correlation between individual differences in modularity and task performance as a function of time between neuroimaging and behavioral testing for the operation span task (green),

the task-shifting task (red), conflict effect of the ANT (purple) and orienting effect of the ANT (blue). Error bars are 95% Confidence Intervals around the correlation values.

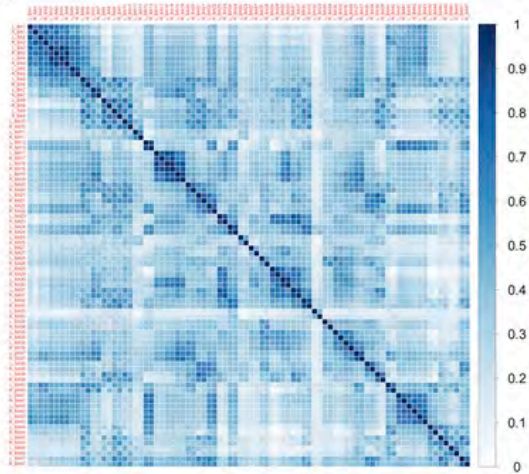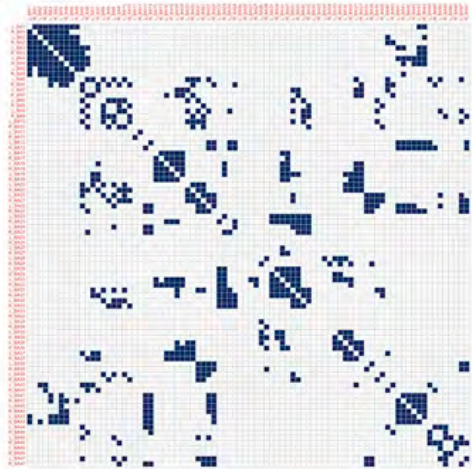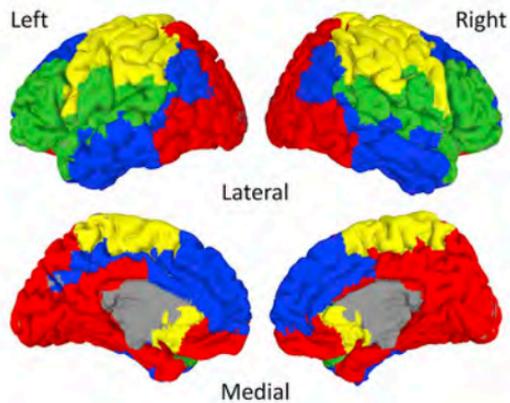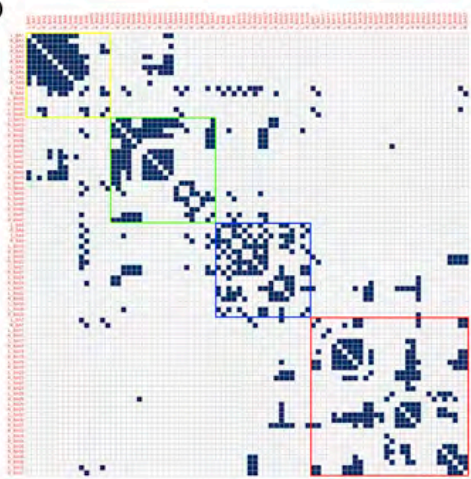

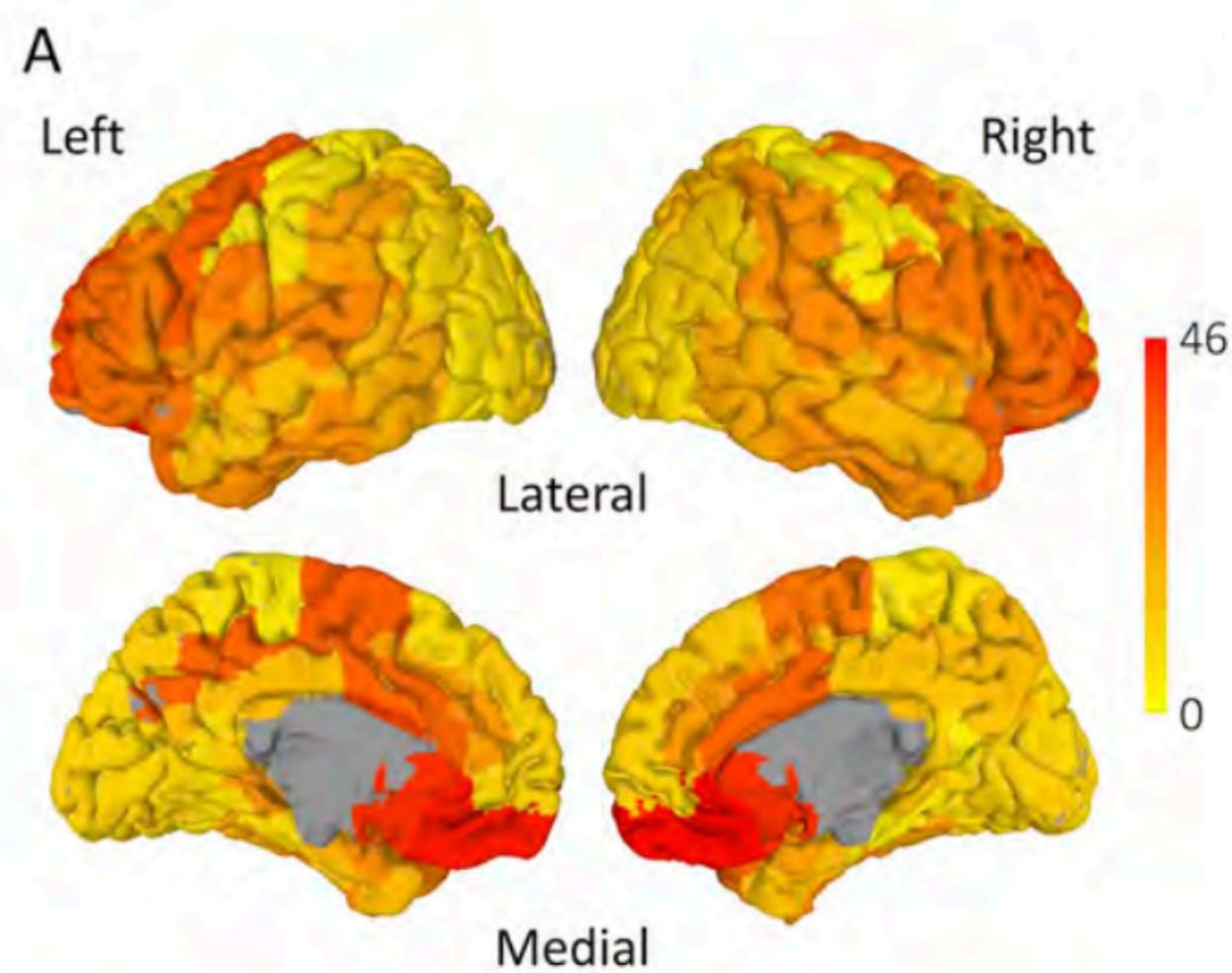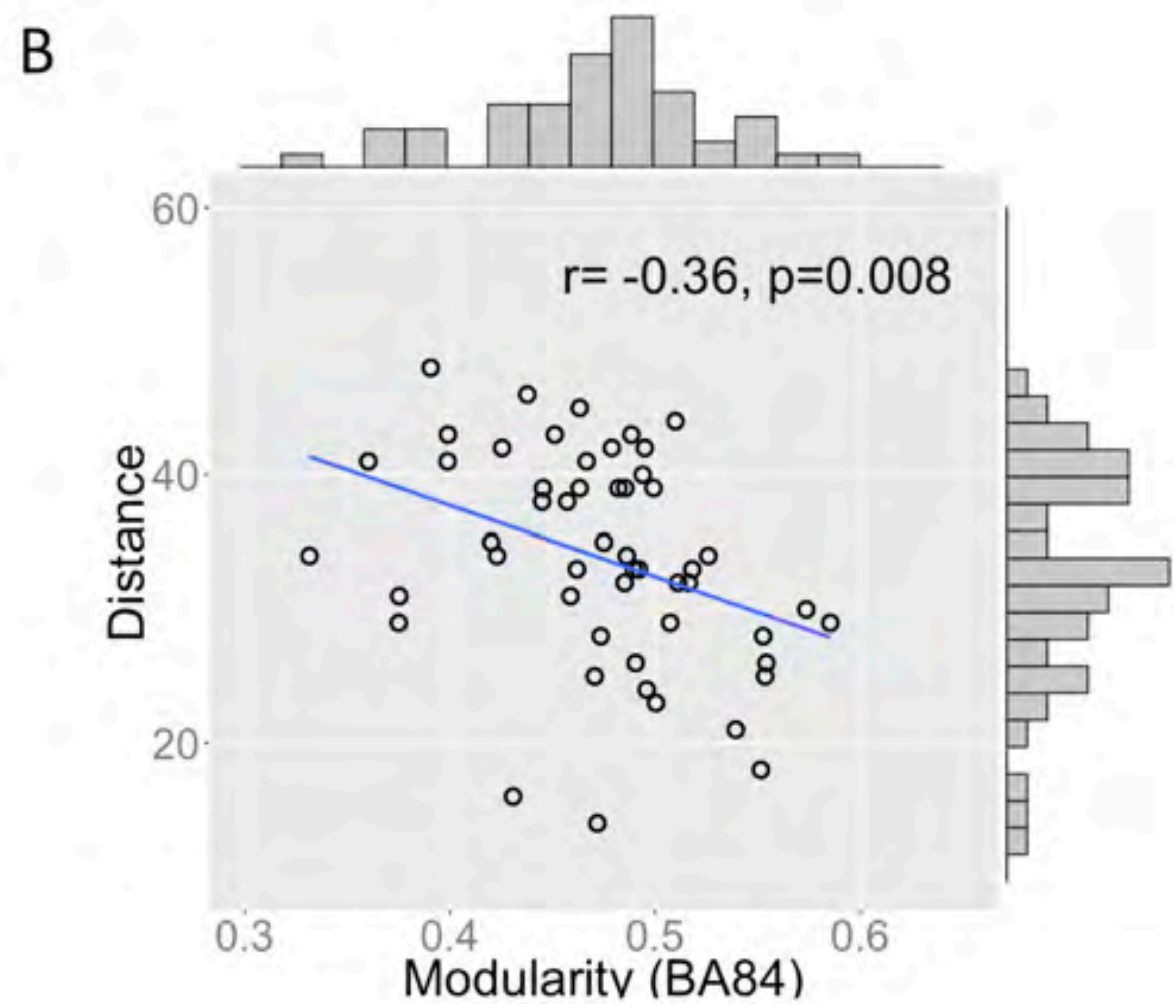

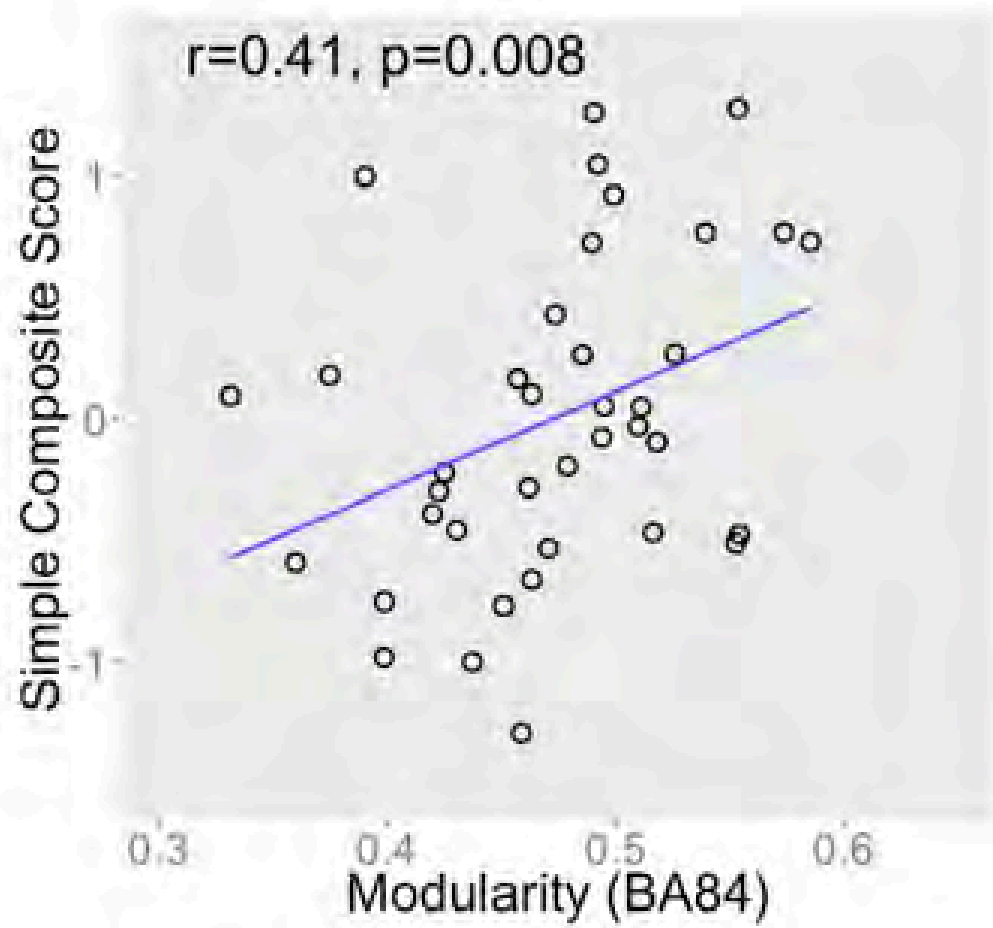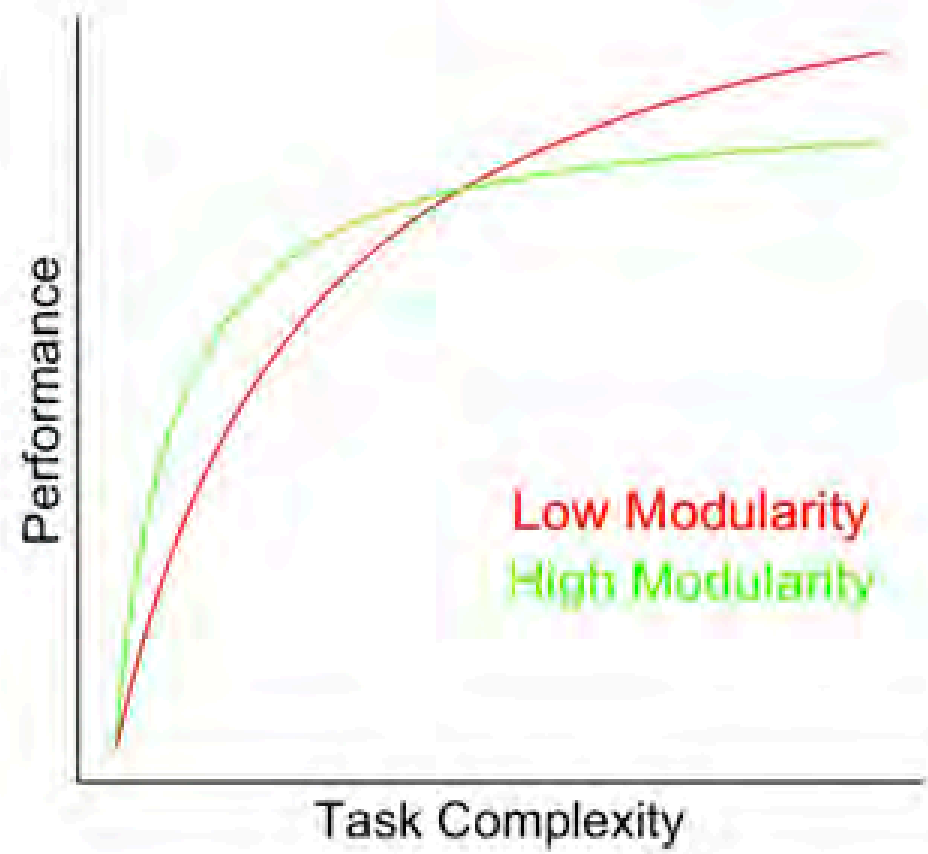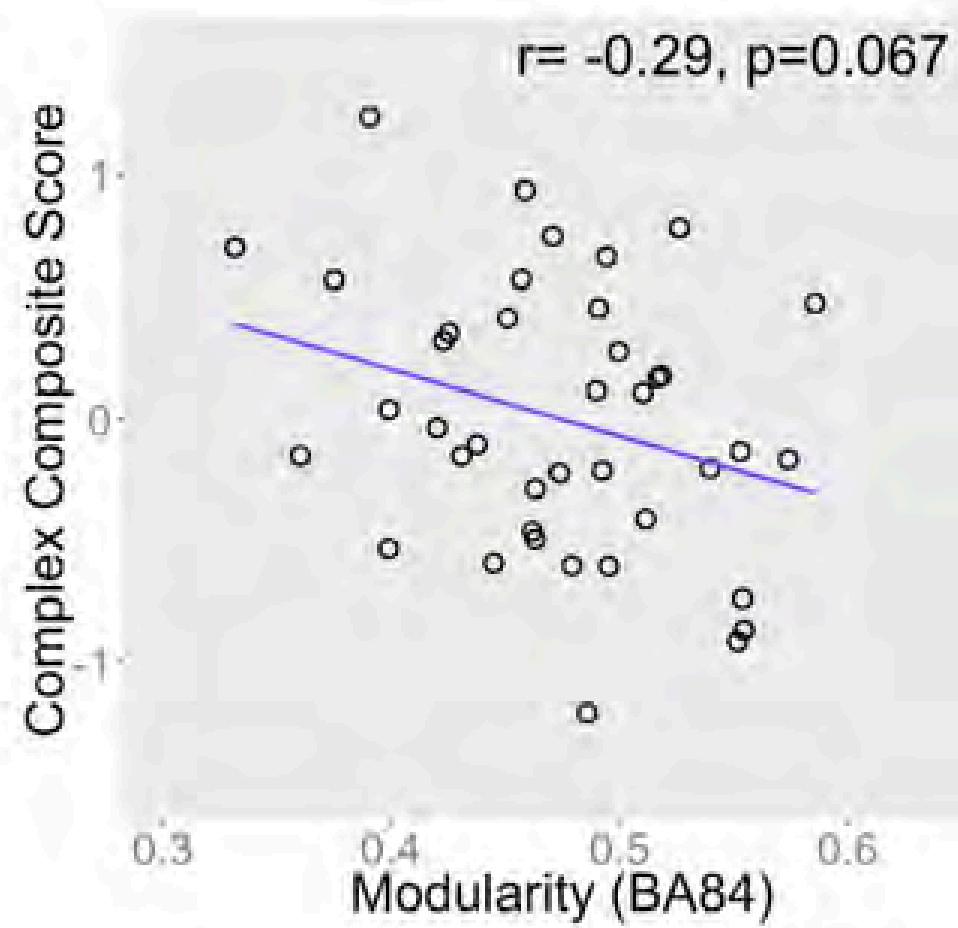

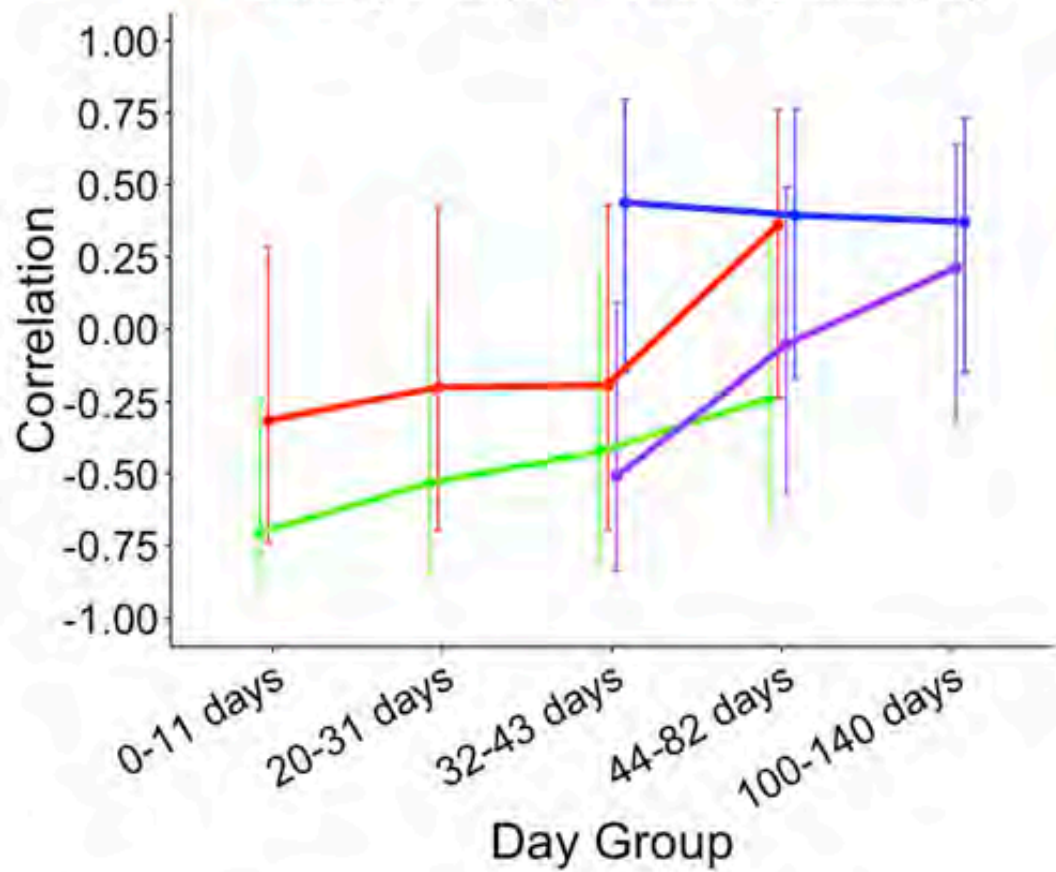